# Design of meta-surface lens integrated with pupil filter[*]

ZHONG Runhui, LING Jinzhong, LI Yangyang, YANG Xudong, WANG Xiaorui

School of Optoelectronic Engineering, Xidian University, Xi'an 710071, China

**Abstract**

Metasurface lenses are miniature flat lenses that can precisely control the phase, amplitude, and polarization of incident light by modulating the parameters of each unit on the substrate. Compared with conventional optical lenses, they have the advantages of small size, light weight, and high integration, and are the core components of photonic chips. Currently, the hot topics for metasurface lens are broadband and achromatic devices, and there is still less attention paid to the resolution improvement. To break through the diffraction limit and further improve the focusing performance and imaging resolution of metasurface lenses, we use unit cells to perform multi-dimensional modulation of the incident light field. Specifically, in this paper, we combine phase modulation of metasurface lens with a pupil filtering, which has been widely applied to traditional microscopy imaging and adaptive optics and has demonstrated powerful resolution enhancement effects. The integrating of these two technologies will continue to improve the imaging performance of metasurface lenses and thus expanding their application fields. In this work, we firstly design a single-cell meta-surface lens composed of a silicon nanofin array and a silica substrate as a benchmark for comparing the performance of integrated meta-surface lens. The lens achieves an ideal focal spot for incident light at 633 nm, resulting in a full width at half maximum (FWHM) of 376.0 nm. Then, a three-zone phase modulating pupil filter is proposed and designed with the same aperture of metasurface lens, which has a phase jump of 0-π-0 from the inside to the outside of the aperture. From the simulation results, the main lobe size of the focal spot is compressed obviously. In the optimization, its structural parameters are scanned for the best performance, and an optimal set of structural parameters is selected and used in the integrated metasurface lens. Finally, the integrated metasurface lens is designed by combining the metasurface lens with a three-zone phase modulating pupil filter, and the FWHM of its focal spot is compressed to 323.4 nm ($\approx 0.51\lambda$), which is not only 15% smaller than original metasurface lens's FWHM of 376.0 nm, but also much smaller than the diffraction limit of $0.61\lambda/NA$ (when $NA = 0.9$, it is approximately 429.0 nm). This result preliminarily demonstrates the super-resolution performance of the integrated meta-surface lens. With the comprehensive

---

regulation of multi-dimensional information, such as amplitude, polarization, and vortex, the integrated meta-surface optical lens will achieve more excellent super-resolution focusing and imaging performance, and will also be widely used in the fields of super-resolution imaging, virtual reality, and three-dimensional optical display, due to its characteristics of high resolution, high integration, and high miniaturization.

**Keywords: meta-lens, pupil filter, super-resolution imaging**

**PACS:** 42.79.-e; 42.40.Lx; 42.25.Bs; 78.67.Pt

## 1. Introduction

Optical metasurface is a kind of artificial two-dimensional material, which can realize the overall modulation of reflected or transmitted light field through the special spatial arrangement and parameter setting [1–3] of periodic micro-unit structure. Because of its excellent ability to control the light-field, metasurface devices play an important role in the phase modulation of optical field, polarization modulation, electromagnetic wave propagation control and so on. For example, Li Shaohe and Li Jiusheng have made frequency-coded meta-surfaces for multi-directional manipulation of terahertz energy radiation [4]; Malek et al.[5] applied metasurface to holographic imaging and achieved higher imaging resolution. In addition, metasurface also has great potential in improving solar energy efficiency [6], multi-band infrared stealth [7] and other fields.

In 2016, Khorasaninejad and Capasso [8] used a periodically distributed two-dimensional micro-element array to control the phase of the light field at each point, fabricated a meta-surface lens that can be compared with a commercial lens, and realized the focusing of the incident plane wave, thus obtaining a miniaturized plane lens with sub-wavelength thickness. Since then, metasurface lenses have attracted extensive attention and in-depth research. Based on the principle of transmission phase, Hu et al. Designed a mid-infrared achromatic double-layer superlens with enhanced efficiency [9]; Xu Ping et al. Designed a long-infrared dual-wavelength confocal superlens [10]; Based on the geometric phase principle, Zhang et al.[11] designed an extended depth-of-focus millimeter-wave superlens. These metasurface lenses well demonstrate the application potential of metasurface devices in micro-nano optics and integrated optics, and also realize broadband or achromatic functions. However, the above metasurface lenses obtain the converged spherical wavefront through phase control, but the focusing performance and imaging resolution of the metasurface lenses are not optimized and improved.

In order to break through the diffraction limit and further improve the focusing performance of the metasurface lens, the pupil filtering super-resolution imaging technology is integrated into the design of the metasurface lens, which can greatly reduce the lateral focal spot size and achieve the effect of super-resolution focusing. Super-resolution pupil filter was first proposed by Di Francia [12] in 1952. The pupil filter proposed by him is composed of multiple concentric rings, each of which has different delay phase and transmittance. Theoretically, the central main lobe of the diffraction spot can be compressed to an infinite size, thus improving the resolution of the optical system. At present, there are still many research groups in the field of pupil filtering super-resolution imaging technology. According to the classification of modulation objects, pupil filters can be divided into amplitude pupil filters [13], phase pupil filters [14] and complex amplitude pupil filters [15]. Among them, the amplitude pupil filter has a low spot intensity due to the partial absorption of light; The complex amplitude pupil filter can adjust the amplitude and phase at the entrance pupil at the same time to achieve better super-resolution effect, but its simulation optimization is difficult and requires more accurate processing accuracy; In contrast, the phase pupil filter has high energy utilization rate, relatively simple design optimization and low processing difficulty, so it is more suitable for the preliminary design of super-resolution system, especially the three-zone phase pupil filter.

In this paper, a three-zone phase pupil filter is combined with a geometric phase meta-surface lens to realize the functions of both, that is, the wavefront of the spherical wave is realized and the function of the phase pupil filter is superimposed, so as to realize lateral super-resolution focusing. Firstly, the (finite-difference time-domain, FDTD) method is used to analyze the phase modulation effect of the structural parameters of the microcell on the transmitted light field, and the appropriate structural parameters are selected to ensure that the microcell can achieve a phase change of $2\pi$ within a rotation angle of 180 °; Correspondingly, constructing the meta-surface lens according to the working wavelength, the expected focal length and the clear aperture, and designing the optical phase required by each position on the meta-surface lens; Finally, the geometric phase method is used to calculate the rotation angles of the micro-units at different positions, and the micro-unit array is arranged on the substrate to complete the design of the meta-surface lens. In addition, according to the aperture of the meta-surface lens, a three-zone phase pupil filter is designed and optimized, and then its structural parameters are applied to the design of the fused meta-surface lens to construct a meta-surface lens with pupil filtering function, so as to further regulate the incident light field. It is found that the fused meta-surface lens has a smaller focal spot size than the original meta-surface lens, which improves the focusing performance and lateral resolution of the meta-surface lens, and can be widely used in various micro-nano optical systems and optical instruments.

## 2. Design of geometric phase metasurface lens

As a comparison target, a single element metasurface lens is designed by using geometric phase to compare the improvement of focusing performance of the metasurface lens with pupil filtering. According to the principle of geometric phase [14], the phase mutation of the optical field can be realized by adjusting the rotation angle of the micro-unit on the metasurface. When circularly polarized light is incident on a microcell with a rotation angle of $\theta$, the Jones matrix of the transmitted light field can be expressed as

$$\begin{aligned} E_{out} &= J(\theta) \cdot E_{in} \\ &= R(-\theta) \cdot \begin{pmatrix} A & 0 \\ 0 & B \end{pmatrix} \cdot R(\theta) \cdot E_{in} \\ &= \begin{pmatrix} \cos\theta & -\sin\theta \\ \sin\theta & \cos\theta \end{pmatrix} \cdot \begin{pmatrix} A & 0 \\ 0 & B \end{pmatrix} \\ &\quad \cdot \begin{pmatrix} \cos\theta & -\sin\theta \\ \sin\theta & \cos\theta \end{pmatrix} \cdot \begin{pmatrix} 1 \\ \pm i \end{pmatrix} \\ &= \frac{A+B}{2} \begin{pmatrix} 1 \\ \pm i \end{pmatrix} + \frac{A-B}{2} e^{\pm 2i\theta} \begin{pmatrix} 1 \\ \mp i \end{pmatrix}, \end{aligned} \quad (1)$$

Where $A$ and $B$ are the reflection or transmission coefficients along the long and short axes of the micro-unit structure, respectively, and the matrix formed by them is the polarization response of the micro-unit, and the rotation direction of the incident circularly polarized light determines the value of the sign in the equation (1). The positive sign is taken when the incident light is left-handed circularly polarized, and the negative sign is taken when the incident light is right-handed circularly polarized. It can be seen from the formula (1) that when the circularly polarized light passes through the metasurface, the outgoing light is composed of two parts, one of which has the same rotation direction as the incident circularly polarized light, and the other has the opposite rotation direction as the incident circularly polarized lights. When $A = B$, the outgoing light does not contain circularly polarized light with opposite handedness, and the Jones matrix corresponds to an isotropic optical element; When $A = -B$, the outgoing light contains only circularly polarized light with opposite handedness, and the absolute value of the phase delay is exactly equal to twice the orientation angle of the optical element. Therefore, the structural parameters of the microcell are modulated such that $A = -B$, so that when the rotation angle of the microcell structure is $\theta$, the circularly polarized light with opposite rotation direction will acquire an additional phase of $\pm 2\theta$.

According to the above principle, a silicon-based nano-brick micro-unit structure of the meta-surface lens is designed in this paper, the substrate material is $SiO_2$, the operating wavelength $\lambda$ is set to 633 nm, the aperture $D$ of the lens is set to 10 μm, and the focal length $f$ of the lens is set to 2 μm. The design of the meta-surface lens can be completed by periodically arranging the micro-units on the substrate in a two-dimensional array and controlling the rotation angle of each micro-unit. The structural diagram and the structural parameters of the micro-units are shown in the Fig.1.

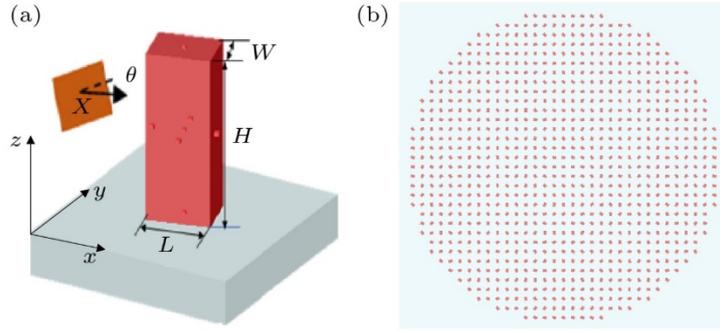

**Figure 1.** Metasurface lens' microcell structure parameters (a) and the periodic arrangement of microcells (b).

In order to realize the phase modulation of the micro-unit to the incident light field, the structural parameters of the micro-unit need to be set to ensure that the phase modulation of $2\pi$ can be realized within the rotation angle range. In addition, it is necessary to avoid the coupling effect between adjacent micro-units of metasurface and the higher diffraction orders of metasurface holography, and to suppress the proportion of the same circular polarization in the transmitted light wave. Considering these two effects, the FDTD algorithm is used to screen and optimize various structural parameters, and finally the period of the micro-cell structure is set to $P$ = 300 nm, the height of the micro-cell is set to $H$ = 400 nm, the length is set to $L$ = 120 nm, and the width is setting to $W$ = 90 nm. With the above parameters, the phase modulation of $2\pi$ can be realized, and the phase delay of the transmitted light field changes with the change of the rotation angle of the micro-unit, as shown in the Fig. 2. The red dotted line is the theoretical curve calculated according to the (1) formula, and the black solid line is the calculation result obtained by using the FDTD algorithm sampling simulation. The two curves are highly consistent and can be used as a reference for the design of the metasurface lens.

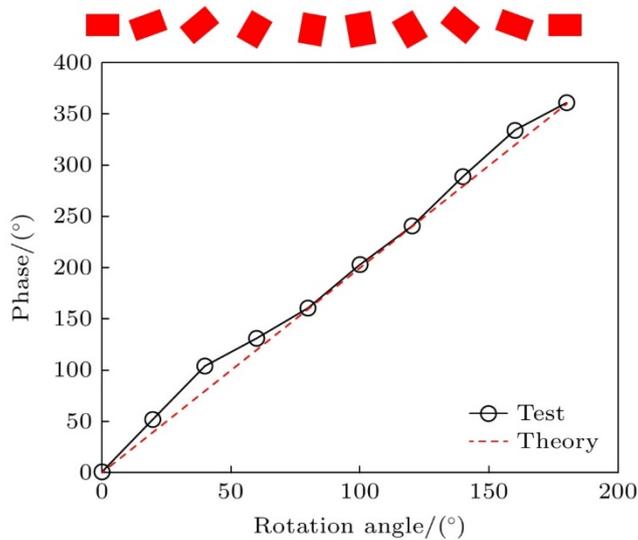

**Figure 2.** Phase changes with the rotation angle of microcell.

According to the working wavelength, aperture and focal length of the metasurface lens, the phase delay required at different positions of the metasurface lens can be calculated to ensure the formation of a convergent spherical wavefront after passing through the meta-surface lens, and different phase delays can be achieved by modulating the rotation angle of the micro-unit at the position. Therefore, the rotation angle $\theta$ of the microcell at the metasurface lens surface coordinate of $(x, y)$ should be set to

$$\theta = (\pi/\lambda)(f - \sqrt{x^2 + y^2 + f^2}), \tag{2}$$

Where $f$ is the focal length of the lens, $\lambda$ is the operating wavelength, and the finished metasurface lens is set as shown in the Fig.1(b). In order to simulate its focusing performance, the metasurface lens is illuminated by right-handed circularly polarized light with a wavelength of 633 nm, and the intensity distribution obtained on the focal plane is shown in Fig.3(a), and the phase distribution is shown in Fig. 3(c); The intensity distribution on the y-z plane behind the lens is shown as Fig. 3(b), and the phase distribution is shown as Fig. 3(d).

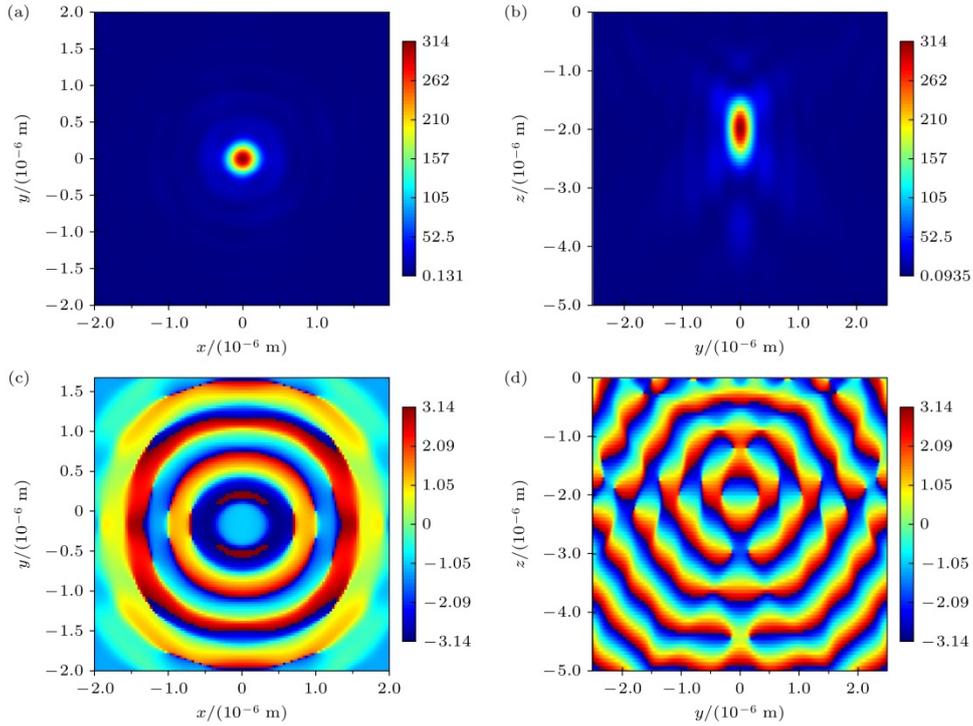

**Figure 3.** (a) Light intensity distributions in *x-y* plane for the focal spot; (b) light intensity distributions in *y-z* plane for the focal spot; (c) phase distributions in *x-y* plane for the focal spot; (d) phase distributions in *y-z* plane for the focal spot.

According to the intensity distribution of the *y-z* plane, the distance between the center of the convergent spot and the meta-surface lens located at $z = 0$ is about 2 μm, which is basically consistent with the design value. The size of the focal spot and the distribution of the side lobes can be seen at the $z = -2$ μm plane, which is the focal plane of the metasurface lens. The (full width at half maximum FWHM) of the focal spot is about 376 nm, and the intensity of the side lobe is not more than 5% of the central intensity, which is an ideal focal spot.

## 3. Design and optimization of pupil filter

According to Born and McCutchen's theory, the amplitude distribution near the focus of an imaging system [16–18] with a pupil function $P(\rho)$ under monochromatic illumination can be expressed as

$$U(\eta, u) = 2\int_0^1 P(\rho)J_0(\eta\rho)\exp(\frac{-ju\rho^2}{2})\rho d\rho, \qquad (3)$$

Where $J_0(\rho)$ is the zero-order Bessel function, $\rho$ is the normalized radius, $\eta$ corresponds to the radial coordinate $r$ on the receiving surface, $r$, $u$ corresponds to the on-axis coordinate $z$ with the focus as the origin, and can be expressed as

$$\eta = kr\sin\alpha, u = kz(\sin\alpha)^2, \qquad (4)$$

Where $k$ is the wave vector of light and $\sin\alpha$ represents the numerical aperture of the system. Through the design and optimization of the pupil function, the size of the focal spot can be reduced to a certain extent, and the effects of super-resolution focusing and super-resolution imaging can be achieved. Generally speaking, the main parameters to evaluate the performance of a super-resolving pupil filter are: the lateral super-resolution factor ($G(T)$), the axial super-resolution factor ($G(A)$), and the Strehl ratio ($S$), which respectively represent the ratio of the central main lobe FWHM of the lateral intensity distribution function near the focus with and without a pupil filter, the ratio of the central main lobe FWHM of the axial intensity distribution function, and the ratio between the maximum central main lobe intensity with and without a filter [19]. In this paper, a three-zone phase pupil filter is designed. Its structure is shown in Fig. 4(a). Its circular aperture is divided into three concentric circles or rings, which are the central zone, the middle zone and the edge zone, and their corresponding normalized radii are $r_1$, $r_2$ and 1, respectively. Fig.4(b) is the optical path of the pupil filter, which is placed in front of the lens to modulate the phase distribution of the incident light field.

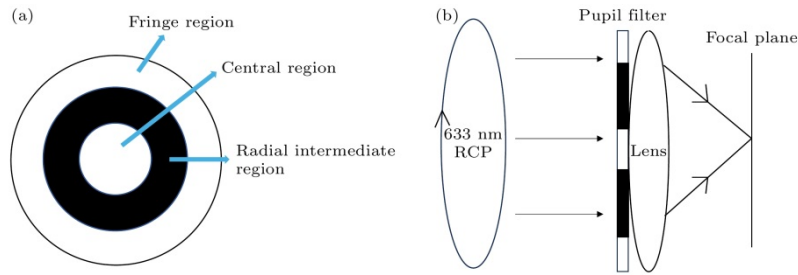

**Figure 4.** Structural diagram (a) and optical path (b) of the three-zone phase pupil filter.

For the above phase pupil filter, the transmittance $t_1 = e^{i0}$ in the central region, the transmittance $t_2 = e^{i\pi}$ in the middle region, and the transmittance $t_3 = e^{i0}$ in the edge region can be set, and the normalized radii of the corresponding regions are $r_1, r_2$, and 1, respectively. Therefore, the lateral super-resolution factor, axial super-resolution factor, and Strehl ratio can be expressed as

$$G(T) = \frac{t_1 r_1^4 + t_2(r_2^4 - r_1^4) + t_3(1 - r_2^4)}{t_1 r_1^2 + t_2(r_2^2 - r_1^2) + t_3(1 - r_2^2)},$$
$$G(A) = 4\left[\frac{t_1 r_1^6 + t_2(r_2^6 - r_1^6) + t_3(1 - r_2^6)}{t_1 r_1^2 + t_2(r_2^2 - r_1^2) + t_3(1 - r_2^2)}\right]$$
$$- 3\left[\frac{t_1 r_1^4 + t_2(r_2^4 - r_1^4) + t_3(1 - r_2^4)}{t_1 r_1^2 + t_2(r_2^2 - r_1^2) + t_3(1 - r_2^2)}\right]^2, \quad (5)$$
$$S = [t_1 r_1^2 + t_2(r_2^2 - r_1^2) + t_3(1 - r_2^2)]^2.$$

In order to improve the lateral resolution of the lens and reduce the energy loss as much as possible, the lateral super-resolution factor and Strehl ratio are taken as the optimization objectives in the simulation optimization process, that is, the larger the $G(T)$ and $S$, the better. The same illumination conditions as the metasurface lens are used in the simulation, that is, the incident light is 633 nm right-handed circularly polarized light; A microlens with the same parameters as the metasurface lens is used to simulate the effect of the pupil filter, that is, the aperture of the microlens is 10 μm and the focal length is 2 μm. Using the FDTD algorithm, the pupil filters with different radius parameters are simulated and analyzed, and the optimal parameters are $r_1 = 0.33$, $r_2 = 0.67$, $r_3 = 1$. Therefore, two other groups of parameters near them are selected for performance comparison, namely, 1) $r_1 = 0.33$, $r_2 = 0.67$, $r_3 = 1$; 2) $r_1 = 0.2$, $r_2 = 0.8$, $r_3 = 1$; 3) $r_1 = 0.1$, $r_2 = 0.9$, $r_3 = 1$. Then, the normalized light intensity distribution after adding the pupil filter was compared with the normalized light intensity distribution of the lens focal spot before adding the pupil filter. The results are shown in Fig. 5.

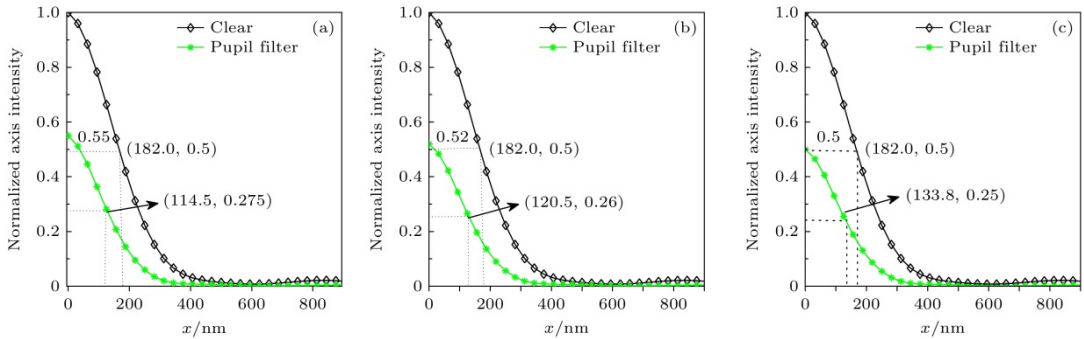

**Figure 5.** Comparison of normalized light intensity distribution: (a) $r_1 = 0.33$, $r_2 = 0.67$, $r_3 = 1$; (b) $r_1 = 0.2$, $r_2 = 0.8$, $r_3 = 1$; (c) $r_1 = 0.1$, $r_2 = 0.9$, $r_3 = 1$.

It can be seen from the Fig. 5 that the pupil filters with the above three structural parameters can improve the lateral resolution, and the corresponding lateral super-resolution factors are 1.59, 1.51 and 1.36, respectively, and the Strehl ratio is maintained at about 0.52. It can be seen that adjusting the width of the middle region of the pupil filter has little effect on the Strehl ratio, but has a significant effect on the lateral super-resolution factor. Therefore, the first group of structural parameters will be used in the subsequent design of the fused meta-surface lens.

## 4. Fusion metasurface lens

Inspired by the working principle of the pupil filter, the focal spot size of the lens can be greatly reduced after the phase of the incident light field is modulated. Therefore, when designing the fused metasurface lens, the phase modulation ability of the micro-unit is used to realize the phase control of the convergent spherical wavefront and the pupil filter at the same time. Specifically, the originally designed metasurface lens is divided into three regions according to the structural parameters of the pupil filter, and different phases are superimposed in different regions to realize the function of the pupil filter, and the phase superposition can be realized by the direction rotation of the micro-unit.

According to the optimization results of pupil filter radius parameters in the previous paper, when the aperture of the lens is 10 μm, the radius of the central area is 1. 65 μm, the radius of the radial middle area is 3. 35 μm, and the diameter of the peripheral area is 5 μm are the optimal parameters. Therefore, the metasurface lens can be divided into three regions, corresponding to the central region, the middle region and the peripheral region of the pupil filter, respectively. The structure is shown as Fig. 6, and different phases are superimposed in different regions, where the peripheral region and the central region keep the phase unchanged, while the middle region increases the phase delay by π. According to the phase modulation characteristic of the metasurface micro-unit, namely, the azimuthal rotation of the micro-unit is $\theta$, the phase of the incident right-handed circularly polarized light changes by $2\theta$. Therefore, rotating each microcell in the middle region by 90 ° gives this region an additional phase delay of π.

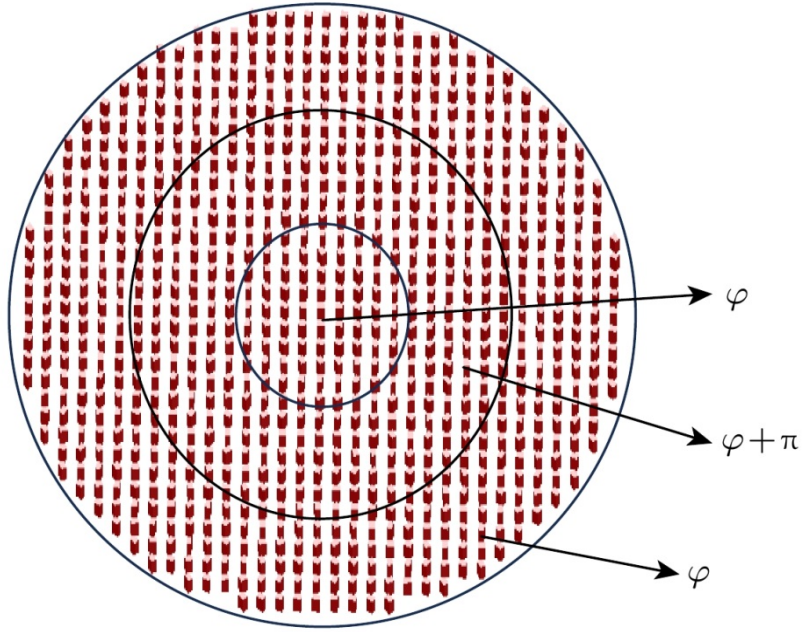

**Figure 6.** Structural diagram of the integrated metasurface lens.

The control and convergence functions of the fused metasurface lens to the incident light field are simulated and analyzed by using the FDTD algorithm. When the right circularly polarized light with the wavelength of 633 nm is incident on the metasurface lens, the phase distribution of the rear *y-z* plane is changed compared with that of the original metasurface lens, as shown in the Fig. 7. Behind the original super-surface lens, a convergent spherical wave is generated, the equi-phase surface distribution of which is relatively regular, and the focus position of which is clearly visible, as shown in the Fig. 7(a); Behind the meta-surface lens with fused pupil filtering, the phase distribution is shown in the Fig. 7(b), and the equi-phase surface is no longer a continuous curved surface, but is divided into several regions with a phase difference of π between adjacent regions, and the focal spot position is moved downward by about 1. 3 µm.

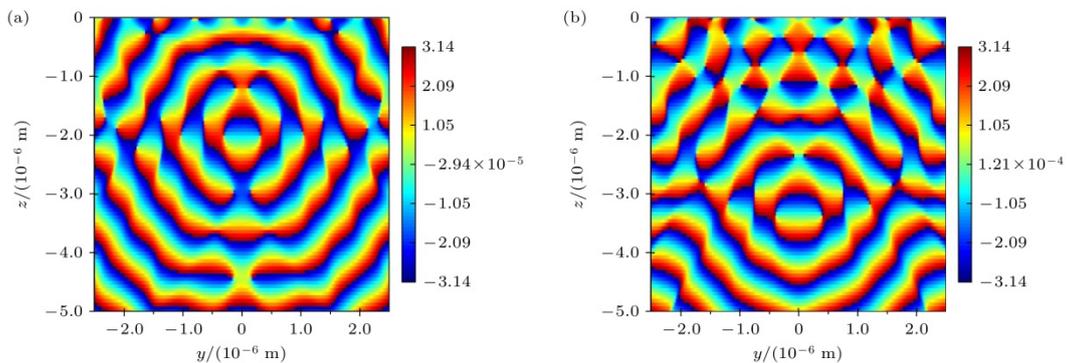

**Figure 7.** (a) Single cell metasurface lens *y-z* plane phase distribution; (b) integrated type metasurface lens *y-z* plane phase distribution

In order to accurately describe the focusing performance of the fused metasurface lens, the intensity distribution on the focal plane and the intensity distribution in the *y-z* plane behind the metasurface lens can be simulated and analyzed, as shown in Fig. 8(a),(b), respectively, and compared with the focal spot characteristics of the original meta-surface lens. From the Fig. 8(a), it can be found that the focal spot size of the fused metasurface lens is significantly smaller than that of the original metasurface lens, and the intensity of its side lobes is not significantly improved, so it is still suitable for wide-field imaging. Compared with the original metasurface lens, the focal spot intensity of the fused metasurface lens is reduced, so higher illumination intensity is required for imaging; The Fig. 8(b) is the light intensity distribution in the *y-z* plane behind the fused metasurface lens, and its focal spot position is displaced from that of the original metasurface lens, and a secondary focus with weaker brightness appears above the focal spot, and the distance between the two centers is more than 2 μm, so the impact on micro-nano processing or super-resolution imaging is very limited.

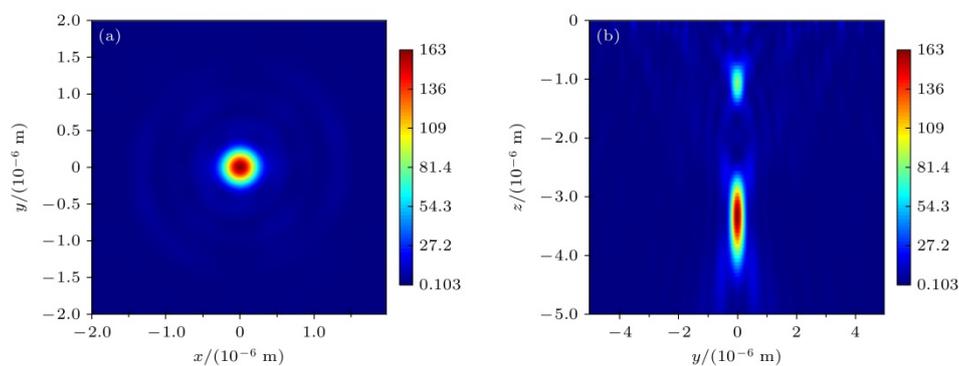

**Figure 8.** Light intensity distribution at the focal spot of integrated metasurface lens: (a) Focal plane image; (b) *y-z* plane.

Although the Fig. 8 intuitively shows the focal spot characteristics and light intensity distribution of the fused metasurface lens, it is impossible to qualitatively analyze the change of the focal spot size. Therefore, the light intensity distribution curve at the center of the focal plane is drawn and compared with the original metasurface lens, which can qualitatively show the performance improvement effect of the fused meta-surface lens.

The Fig. 9 is the normalized intensity distribution curve at the focal plane center of the two kinds of metasurface lenses, where the black curve represents the original metasurface lens and the green curve represents the meta-surface lens with pupil filtering. By comparing the two curves in the figure, it can be seen that the FWHM of the original metasurface lens is about 376 nm, while the FWHM of the fused pupil filtered metasurface lens is about 323.4 nm. It can be seen that the transverse width of the main lobe of the focal spot of the fusion metasurface is compressed to a certain extent, the transverse super-resolution factor is calculated to be about 1.29, and the transverse resolution is improved by about 15%, but its central brightness is reduced, and its Strehl ratio is about 0.51. With the continuous compression of the focal spot size, the intensity of its side lobe will continue to increase, and the central brightness will continue to decrease, so it is necessary to balance the parameters reasonably and compress as much as possible without affecting the imaging quality.

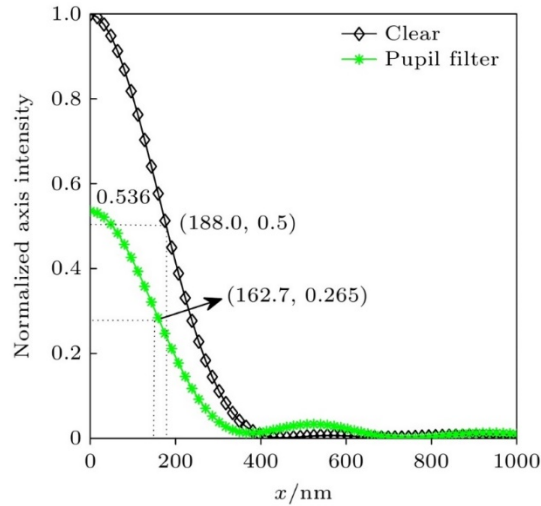

**Figure 9.** Comparison of normalized light intensity distributions of two metasurface lenses around the focal spot centrals.

## 5. Conclusion

In this paper, a single cell meta-surface lens composed of nano-silicon array and $SiO_2$ substrate is designed to achieve an ideal focusing spot for 633 nm incident light. Then, a three-zone phase pupil filter is proposed based on the simulation design of the pupil filter with the same aperture, and the phase mutation from inside to outside is 0-π-0, so that the size of the main lobe of the focal spot is compressed, and the pupil filters with different radius parameters are simulated and optimized. Finally, a novel fused metasurface lens is obtained by fusing the above metasurface lens with a three-zone phase pupil filter, and the FWHM of its focal spot is compressed to 323.4 nm (≈ 0.51λ), which is not only greatly improved compared with the 376.0 nm of a common meta-surface lens, but also exceeds the diffraction-limited 0.61λ/NA (when NA = 0.9, about 429.0 nm), and this result preliminarily demonstrates the super-resolution effect of the fused meta-surface lens. With the comprehensive control of multi-dimensional information such as amplitude, polarization and vortex, the fused metasurface optical lens will achieve more excellent super-resolution focusing and imaging performance, and will be widely used in super-resolution imaging, virtual reality, three-dimensional optical display [20,21] and other related fields with its high-resolution, high-integration and miniaturization characteristics.